\documentclass[aps,
twocolumn,
superscriptaddress]{revtex4}%
\usepackage{amsfonts}
\usepackage{amsmath}
\usepackage{amssymb}
\usepackage{graphicx}%
\usepackage[dvipsnames]{xcolor}
\newcommand{\beginsupplement}{%
        \setcounter{table}{0}
        \renewcommand{\thetable}{S\arabic{table}}%
        \setcounter{figure}{0}
        \renewcommand{\thefigure}{S\arabic{figure}}%
}

\begin{document}

\title{Unusual Mott transition associated with charge-order melting in BiNiO$_3$ \\ under pressure}

\author{I. Leonov}
\affiliation{M. N. Miheev Institute of Metal Physics, Russian Academy of Sciences, 620108 Yekaterinburg, Russia}
\affiliation{Materials Modeling and Development Laboratory, National University of Science and Technology 'MISiS', 119049 Moscow, Russia}

\author{A. S. Belozerov}
\affiliation{M. N. Miheev Institute of Metal Physics, Russian Academy of Sciences, 620108 Yekaterinburg, Russia}
\affiliation{Ural Federal University, 620002 Yekaterinburg, Russia}

\author{S. L. Skornyakov}
\affiliation{M. N. Miheev Institute of Metal Physics, Russian Academy of Sciences, 620108 Yekaterinburg, Russia}
\affiliation{Ural Federal University, 620002 Yekaterinburg, Russia}

\begin{abstract}
We study the electronic structure, magnetic state, and phase stability of paramagnetic BiNiO$_3$ near a pressure-induced Mott insulator-to-metal transition (MIT) by employing a combination of density functional and dynamical mean-field theory. We obtain that BiNiO$_3$ exhibits an anomalous negative-charge-transfer insulating state, characterized by charge disproportionation of the Bi $6s$ states, with Ni$^{2+}$ ions. Upon a compression of the lattice volume by $\sim$4.8\%, BiNiO$_3$ is found to make a Mott MIT, accompanied by the change of crystal structure from triclinic $P\bar{1}$ to orthorhombic $Pbnm$. The pressure-induced MIT is associated with the melting of charge disproportionation of the Bi ions, caused by a charge transfer between the Bi $6s$ and O $2p$ states. The Ni sites remain to be Ni$^{2+}$ across the MIT, which is incompatible with the valence-skipping Ni$^{2+}$/Ni$^{3+}$ model. Our results suggest that the pressure-induced change of the crystal structure drives the MIT in BiNiO$_3$.  
\end{abstract}

\maketitle


The Mott metal-insulator transition driven by correlation effects has been an outstanding problem in condensed matter physics over many decades~\cite{Imada}. 
In recent years, increasing attention has been drawn to the rare earth nickelate perovskites $R$NiO$_3$ ($R$ = rare earth, $R^{3+}$) with a high oxidation state of nickel, Ni$^{3+}$ $3d^7$~\cite{Nickelates,Subedi_2015,Mercy2017}. $R$NiO$_3$ compounds (except for LaNiO$_3$) exhibit a sharp metal-insulator transition (MIT) upon cooling below $T_\mathrm{MIT}$~\cite{Torrance1992}.
The phase transition is accompanied by a structural transformation from an orthorhombic ($Pbnm$, GdFeO$_3$-type) to monoclinic ($P2_1/n$) crystal structure, with a cooperative breathing distortion of NiO$_6$ octahedra~\cite{Torrance1992}. 

Based on the Ni-O bond lengths analysis and X-ray absorption spectroscopy, a partial $ \mathrm{Ni}^{(3\pm \delta)+}$ charge disproportionation of Ni ions was proposed to occur in the insulating $R$NiO$_3$ phases~\cite{Torrance1992,Medarde2009}.
By contrast, further electronic structure calculations explain the insulating state of $R$NiO$_3$ in terms of bond disproportionation, with alternating Ni ions which (nearly) adopt a Ni$^{2+}$ $3d^8$ (Ni$^{2+}$ ions with local moments) and $3d^8\underline{L}^2$ (nonmagnetic spin-singlet) electronic configuration ($\underline{L}$ denotes a hole in the O $2p$ band)~\cite{Park2012,Johnston2014,Subedi_2015}. 
The transition temperature $T_\mathrm{MIT}$ is strongly related to the degree of structural distortion of $R$NiO$_3$, determined by the size of $R$-ions. With decrease of the $R$-ionic radius, the Ni-O-Ni bond angle, which determines the degree of overlapping of the Ni $3d$ and O $2p$ orbitals (and hence the Ni $3d$ bandwidth), becomes smaller and $T_\mathrm{MIT}$ is increased. In accord with this, the least distorted LaNiO$_3$ is found to be a correlated metal~\cite{Torrance1992,Nowadnick2015}. 
In this context, the replacement of La$^{3+}$ with a larger ion, such as Bi$^{3+}$, should in principle result in a metal with (nearly) cubic perovskite structure.
By contrast, BiNiO$_3$ has been found to be an insulator with a highly distorted perovskite structure (triclinic, $P\bar{1}$) and unusual valence ordering of the $A$-site Bi ions~\cite{Ishiwata2002}. In particular, based on X-ray and neutron diffraction, it was proposed that Ni ions adopt a Ni$^{2+}$ state, with an electronic configuration Bi$^{3+}_{0.5}$Bi$^{5+}_{0.5}$Ni$^{2+}$O$_3$~\cite{Ishiwata2002,Azuma2007,Carlsson2008}.


BiNiO$_3$ is known due to its colossal negative thermal expansion across the pressure-induced MIT, as suggested caused by a Bi/Ni charge transfer~\cite{Azuma2011}.
Under ambient conditions, BiNiO$_3$ crystallizes in a triclinic perovskite crystal structure (space group $P\bar{1}$, a subgroup of $P2_1/n$) with two inequivalent Bi and four Ni sites~\cite{Azuma2007} (see Supplementary Fig. S1~\cite{supplement} and Ref.~\cite{vesta} therein). It is an insulator with an energy gap of 0.68~eV~\cite{Ishiwata2002}. Below the N\'eel temperature of ${T_\mathrm{N}\sim 300}$~K, BiNiO$_3$ is a $G$-type antiferromagnet with a near-antiferromagnetic alignment of Ni$^{2+}$ ${S = 1}$ spins, implying a predominant role of the antiferromagnetic Ni-O-Ni superexchange~\cite{Ishiwata2002,Carlsson2008,footnote1}.
Moreover, similarly to the small $R$-ions $R$NiO$_3$ the (charge-disproportionated) paramagnetic insulating phase of BiNiO$_3$ extends well above $T_\mathrm{N}$, implying the crucial importance of correlation effects~\cite{footnote2,Park2012,Subedi_2015}. BiNiO$_3$ shows a Mott insulator-to-metal phase transition (in the paramagnetic phase) under pressure (above $\sim$4 GPa) or upon substitution of the $A$-site Bi ions with La~\cite{Ishiwata2005,Wadati2005}. In close similarity to $R$NiO$_3$, the MIT is accompanied by the change of crystal structure from the triclinic $P\bar{1}$ (insulating) to orthorhombic GdFeO$_3$-type $Pbnm$ (metallic) phase, with a volume collapse of $\sim$3\% and melting of charge disproportionation (Ni and Bi sites are equivalent in the $Pbnm$ structure of BiNiO$_3$). Based on the powder X-ray absorption and neutron diffraction, it was proposed that the melting of charge disproportionation leads to a charge transfer from Ni$^{2+}$ to Bi$^{3+}$, so that the electronic state of the $Pbnm$ metallic phase can be described as Bi$^{3+}$Ni$^{3+}$O$_3$~\cite{Azuma2007,Mizumaki2009}. This valence distribution however is in odd with photoemission spectroscopy results for $Pbnm$ BiNiO$_3$ that reveal that the nickel valence is far from being Ni$^{3+}$~\cite{Wadati2005}.

The electronic properties of BiNiO$_3$ have recently been calculated using band-structure methods supplemented with the on-site Coulomb correlations for the Ni $3d$ states within density-functional theory (DFT)+$U$~\cite{dft_plus_u} and dynamical mean-field theory (DMFT)~\cite{dmft} methods~\cite{Cai2007}. However, these studies have mostly been focused on the valence skipping model, with a valence transition between the charge-ordered insulating [Bi$^{3+}_{0.5}$Bi$^{5+}_{0.5}$][Ni$^{2+}$] and the uniform metallic [Bi$^{3+}$][Ni$^{3+}$] state, assuming a long-range magnetic ordering. 
In fact, however, the MIT transition in BiNiO$_3$ is known to occur in the \emph{paramagnetic} state, implying the importance of electronic correlations. 
Moreover, a recent electronic structure study of BiNiO$_3$ using DFT and slave rotor methods suggests that BiNiO$_3$ is a self-doped Mott insulator~\cite{Saha_Dasgupta}.

In this paper, we explore the evolution of the electronic structure, magnetic state, and phase stability of paramagnetic BiNiO$_3$ near the pressure-induced Mott MIT.
We employ a fully self-consistent in charge density DFT+DMFT approach~\cite{dftdmft} 
implemented with plane-wave pseudopotentials~\cite{espresso,Leonov1}
which makes it possible to capture all generic aspects of the interplay between the electronic correlations, magnetic states, and crystal structure of BiNiO$_3$ near the Mott MIT \cite{dftdmft_aplications}. 
The DFT+DMFT calculations explicitly include the Bi $6s$, O $2p$, and Ni $3d$ valence states, by constructing a basis set of atomic-centered Wannier functions within the energy window spanned by the $s$-$p$-$d$ band complex~\cite{Wannier}. 
This allows us to take into account a charge transfer between the Bi $6s$, O $2p$, and Ni $3d$ states, accompanied by the strong on-site Coulomb correlations of the Ni $3d$ electrons.
We use the continuous-time hybridization-expansion (segment) quantum Monte-Carlo algorithm in order to solve the realistic many-body problem~\cite{CT-QMC}. We take the average Hubbard ${U=6}$~eV and Hund's exchange ${J=0.95}$~eV as estimated previously for $R$NiO$_3$ \cite{Park2012,Nowadnick2015}. We use the fully localized double-counting correction, evaluated from the self-consistently determined local occupations, to account for the electronic interactions already described by DFT.

\begin{figure}[t]
\includegraphics[trim=0cm 0cm 0cm 0cm,width=0.36\textwidth]{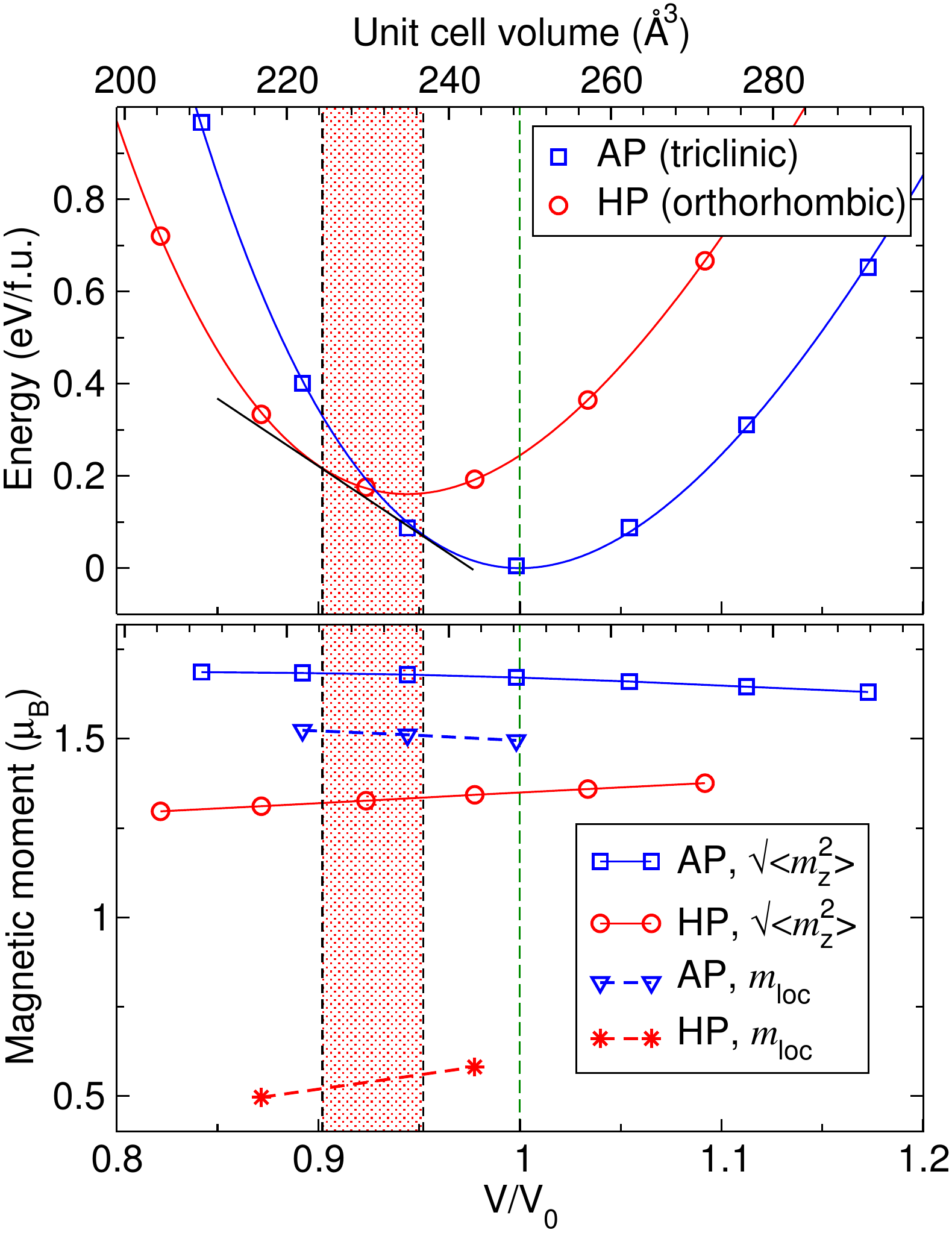}
\caption{(Color online) Total energy (top) and local magnetic moments (bottom) of paramagnetic BiNiO$_3$ obtained by DFT+DMFT for the ambient-pressure $P\bar{1}$ (AP) and high-pressure $Pbnm$ (HP) phases as a function of the unit cell volume at a temperature ${T = 387}$~K.}
\label{fig:energy_and_moments}
\end{figure}

In Fig.~\ref{fig:energy_and_moments} we display our DFT+DMFT results for the phase equilibrium and local magnetic moments of Ni ions of paramagnetic BiNiO$_3$.
In these calculations, we adopt the crystal structure data for the ambient pressure triclinic $P\bar{1}$ and high-pressure orthorhombic $Pbnm$ structures (taken at a pressure of $\sim$7.7~GPa) from experiment~\cite{Azuma2007}, and evaluate the DFT+DMFT total energies as a function of lattice volume. 
Overall, our results for the electronic structure and lattice properties
of BiNiO$_3$ agree well with experimental data~\cite{Wadati2005,Ishiwata2002,Azuma2007,Carlsson2008,Azuma2011}.
In particular, the triclinic $P\bar{1}$ phase is found to be thermodynamically stable at ambient pressure, with a total-energy difference between the ambient-pressure and high-pressure phases of $\sim$160 meV/f.u.. The calculated equilibrium lattice volume $V_0=248.8~\AA^3$ and bulk modulus $K_0=149$~GPa ($K'\equiv dK/dP$ is fixed to $K'=4$). 
Interestingly, all the Ni sites (the insulating $P\bar{1}$ phase has four inequivalent Ni sites) are nearly equivalent and are in the Ni$^{2+}$ state. 
The Ni$^{2+}$ state is also confirmed by the eigenvalues analysis of the reduced Ni $3d$ density matrix, which suggests that the Ni ions are in the $\sqrt{0.63} |d^8 \rangle + \sqrt{0.32} | d^9 \rangle$ state (all the rest contributions are below 0.05).
Moreover, the calculated local (instantaneous) magnetic moment $\sqrt{\langle \hat{m}^2_z \rangle} \simeq 1.67\,\mu_\textrm{B}$, agrees with the high-spin ${S=1}$ state of the Ni$^{2+}$ ions.

\begin{figure}[h]
\includegraphics[trim=0cm 0cm 0cm 0cm,width=0.48\textwidth]{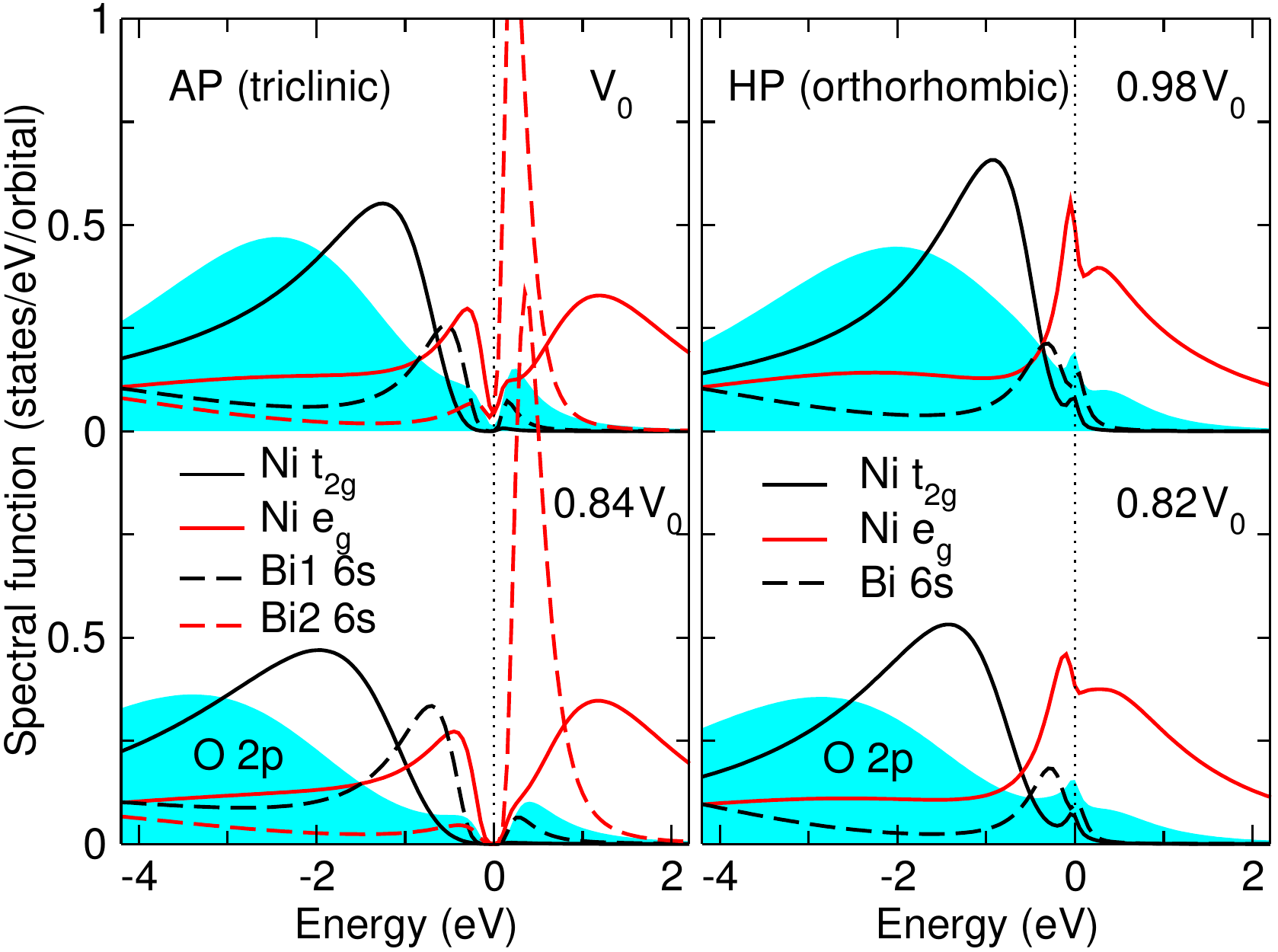}
\caption{(Color online)
Orbitally-resolved spectral functions of BiNiO$_3$ calculated within DFT+DMFT using the maximum entropy method for the ambient-pressure $P\bar{1}$ (left panel) and high-pressure $Pbnm$ (right panel) structures for different unit cell volumes at a temperature ${T = 387}$~K. 
\label{fig:dos}}
\end{figure}

Our calculations for the insulating $P\bar{1}$ phase of BiNiO$_3$ give a self-doped Mott insulator \cite{self-doped} with an energy gap of $\sim$0.3~eV (see the left panel of Fig.~\ref{fig:dos}), in agreement with the resistivity and photoemission experiments \cite{Wadati2005,Ishiwata2002} (see also Supplementary Fig. S3).
In particular, the energy gap lies between the occupied and unoccupied Ni $e_g$ states, strongly mixed with the O $2p$ and the empty Bi2 $6s$ states (the Bi1 $6s$ states are fully occupied).
The O $2p$ states are about -3.6~eV below the Fermi level, but have a substantial contribution both above and below $E_\textrm{F}$. The latter is due to the strongly covalent B $6s$--O $2p$ bonding, suggesting creation of a ligand hole caused by a charge transfer between Bi $6s$ and O $2p$.
While the occupied Bi1 and Bi2 $6s$ states are seen to be localized deep below $E_\textrm{F}$, at about -10~eV, the empty Bi2 $6s$ states appear right at the bottom of the conduction band, with a sharp resonant peak at $\sim$0.4~eV.
The top of the valence band has a mixed Ni $3d$ and O $2p$ character, with a resonant peak in the filled $e_g$ bands located at about -0.4 eV below the Fermi level, which can be ascribed to the formation of a Zhang-Rice bound state \cite{Zhang1988}.
%

Our result for the insulating $P\bar{1}$ phase is characterized by a remarkable charge disproportionation of the Bi $6s$ states (due to the appearance of two different Bi sites with sufficiently different oxygen environment in the insulating phase). In fact, while the Bi1 $6s$ states are almost completely occupied, the Bi2 $6s$ Wannier occupancy is only about 1.56. This implies a charge difference of ${\Delta N_{\mathrm{Bi}\text{-}6s} \sim 0.42}$, i.e., it is about 21\% of the ideal Bi$^{3+}$-Bi$^{5+}$ value. 
Interestingly, the corresponding Bi $6s$ charge difference is in agreement with a charge disproportionation of $\sim$0.2 (i.e., of $\sim$20\% of the ideal valence skipping) found in the low-temperature charge-ordered phases of the mixed-valent oxides, such as Fe$_3$O$_4$~\cite{mixed-valent-oxides}, and of $\sim$0.2-0.3 charge disproportionation of the Ni ions in $R$NiO$_3$~\cite{Torrance1992}. Moreover, previous estimates for the bond-disproportionated insulating phases of the bismuth perovskites BaBiO$_3$ and SrBiO$_3$ show a small charge disproportionation between the Bi ions of $\sim$0.3~\cite{Foyevtsova}.
We also verified our result for $\Delta N_{\mathrm{Bi}\text{-}6s}$ by calculating the corresponding charge difference within the Bi-ion radius of 1.31 \AA, a typical value for the Bi$^{3+}$ ion. Nevertheless, we find that the result is 
robust, with ${\Delta N_{\mathrm{Bi}\text{-}6s} \sim 0.34}$. While all the Ni's are in the Ni$^{2+}$ state (and, as we will show below, the Ni$^{2+}$ state remains stable above the MIT in the metallic $Pbnm$ phase) this suggests the stabilization of the charge disproportionated $\mathrm{Bi1}^{3+}_{0.5}(\mathrm{Bi2}^{(3 + \delta)+} \underline{L}^{2-\delta})_{0.5}$ valence configuration in the insulating $P\bar{1}$ phase of BiNiO$_3$. 
We argue that the obtained valence configuration can be rationalized as being intermediate between the two limits: the pure valence skipping Bi$^{3+}$-Bi$^{5+}$ and the Bi-O bond disproportionation $\mathrm{Bi}^{3+}$-[$\mathrm{Bi}^{3+} \underline{L}^{2}$] models.

\begin{figure}[t]
\begin{tabular}{c}
\includegraphics[trim=0cm 0cm 0cm 0cm,width=0.36\textwidth]{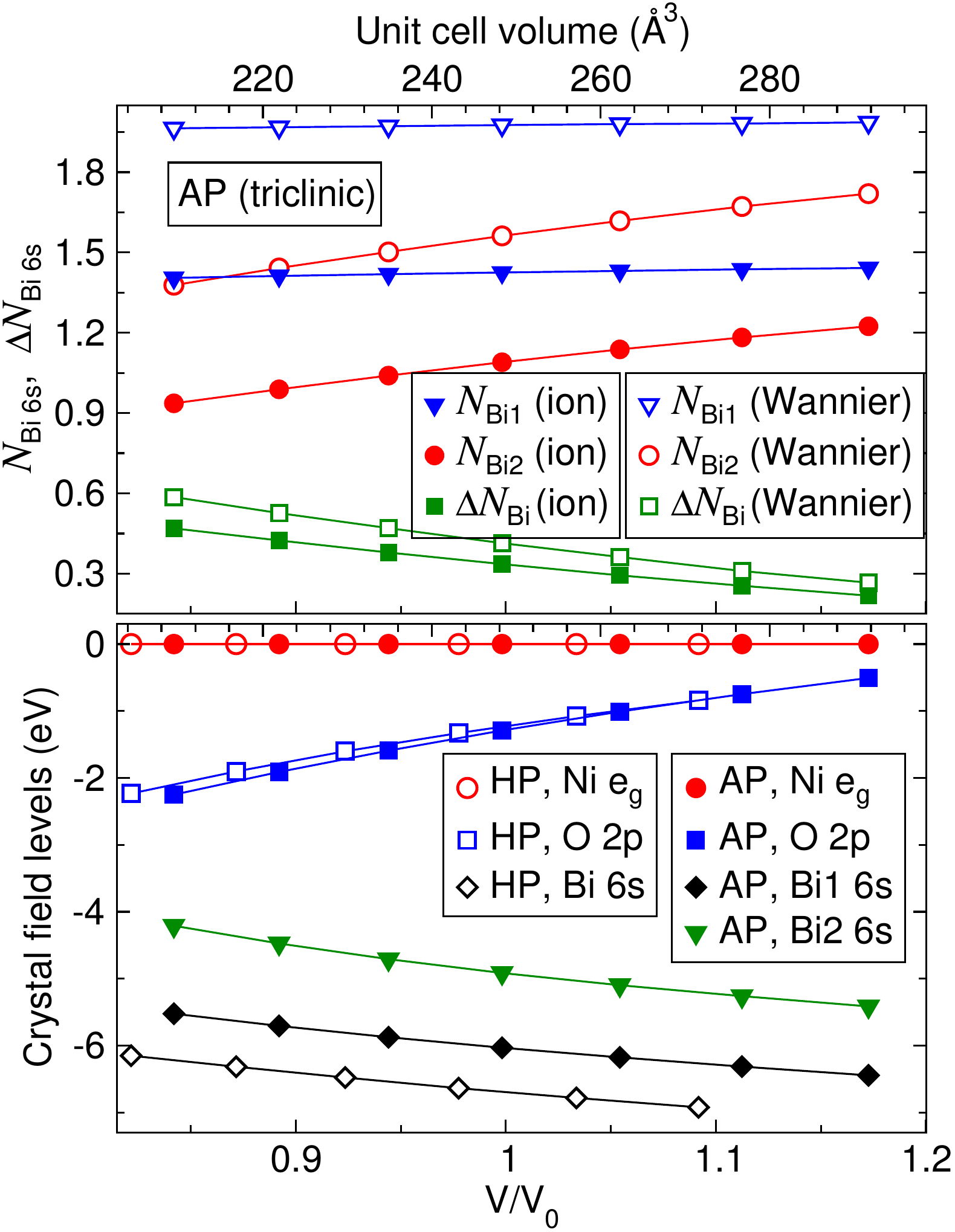}
\end{tabular}
\caption{(Color online)
Bi $6s$ occupations for the ambient-pressure (AP) $P\bar{1}$ phase of BiNiO$_3$ calculated by DFT+DMFT as a function of lattice volume (top). ${\Delta N_\textrm{Bi}}$ denotes the corresponding Bi $6s$ charge disproportionation (${\Delta N_\textrm{Bi}=N_\textrm{Bi1}-N_\textrm{Bi2}}$).
Bottom: the Ni $e_g$ and O $2p$ crystal field levels for the AP and high-pressure $Pbnm$ BiNiO$_3$~\cite{footnote3}. 
\label{fig:occup}}
\end{figure}

Interestingly, the energy gap of the triclinic BiNiO$_3$ phase is seen to increase upon (an uniform) compression (while decreasing and even closing upon expansion) of the unit cell volume (see the lower panel of Fig.~\ref{fig:dos}).
This counter-intuitive change of the energy gap value in a Mott insulator is accompanied by a remarkable increase of charge disproportionation of the Bi ions (under pressure), suggesting the importance of a Bi $6s$-O $2p$ charge transfer.
In particular, our results show that the Bi $6s$ charge disproportionation becomes larger in the $P\bar{1}$ crystal structure of BiNiO$_3$ upon decrease of the lattice volume (see Fig.~\ref{fig:occup}).
Upon compression, the Bi2 $6s$ orbital occupation gradually decreases, whereas the Bi1 $6s$ states are fully occupied, with a nearly constant occupation $N_{\mathrm{Bi1}\text{-}6s} \sim 1.97$. %
In addition, our DFT+DMFT calculations using different Hubbard $U$ values (${U=5}$~eV and 8~eV) show that the energy gap increases upon increasing of $U$, in agreement with the behavior of a Mott insulator. Interestingly, the Bi $6s$ charge disproportionation becomes larger for the larger $U$ values, by $\sim$5\% upon increasing of the $U$ value from ${U=6}$~eV to 8~eV.

This behavior is consistent with the change of the crystal field levels of the Ni $e_g$, O $2p$, and Bi $6s$ states under pressure (see Fig.~\ref{fig:occup}). In fact, the O $2p$ levels are found to shift deep below the Ni $e_g$ states under pressure, whereas the Bi $6s$ states go up in energy. The change of the O $2p$ and Bi $6s$ crystal field levels leads to the enhancement of the Bi $6s$-O $2p$ hybridization under pressure, supporting the hybridization-switching mechanism proposed by Paul \emph{et al.}~\cite{Saha_Dasgupta}.
Our results suggest that the $P\bar{1}$-structured BiNiO$_3$ is an \emph{unconventional} Mott insulator in which the correlated insulating state is in much respect controlled by an $s$-$p$ level splitting between the uncorrelated $A$-site Bi $6s$ and ligand O $2p$ states.

\begin{figure}[t]
\includegraphics[trim=0cm 0cm 0cm 0cm,width=0.39\textwidth]{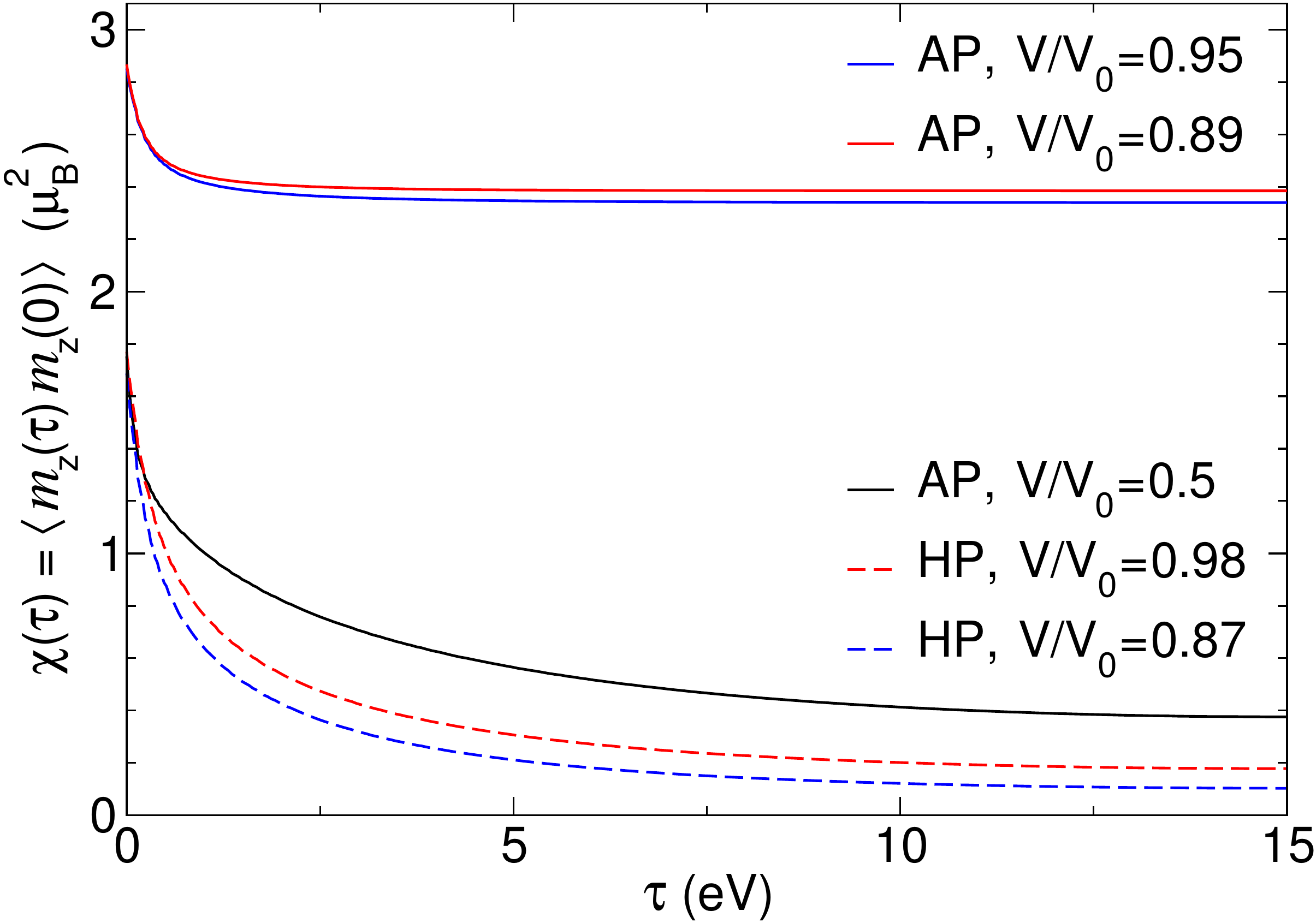}
\caption{(Color online)
Local spin correlation function $\chi(\tau)=\langle \hat{m}_z(\tau) \hat{m}_z(0) \rangle$ of the Ni $3d$ states calculated by DFT+DMFT for the ambient-pressure $P\bar{1}$ (AP) and the high-pressure $Pbnm$ (HP) structures of BiNiO$_3$ for different 
volumes.
\label{fig:chi_loc}}
\end{figure}

Upon further compression the $P\bar{1}$-structured BiNiO$_3$ becomes metallic below ${\sim 0.5\,V_0}$, with the (instantaneous) local moment of $\sim 1.36\,\mu_\textrm{B}$. The MIT is accompanied with a collapse of local moments due to delocalization of the Ni $3d$ electrons, as seen from the behavior of local spin susceptibility $\chi(\tau)=\langle \hat{m}_z(\tau) \hat{m}_z(0) \rangle$ (see Fig.~\ref{fig:chi_loc}). In fact, $\chi(\tau)$ is seen to decay fast with the imaginary time~$\tau$. In agreement with this, the fluctuating moment is only of $\sim$0.75~$\mu_\textrm{B}$ (evaluated as $m_\textrm{loc}=[T \int_0^{1/T} \chi( \tau )d\tau]^{1/2}$), that differs sufficiently from the instantaneous moment. While the Bi $6s$ charge disproportionation is large in the highly-compressed metallic $P\bar{1}$ phase, ${\Delta N_{\mathrm{Bi}\text{-}6s} \sim 1.04}$, this suggests that the Bi~$6s$ charge ordering alone cannot explain the insulating state of BiNiO$_3$. 
In agreement with this, our results for structural optimization of the $P\bar{1}$ phase within nonmagnetic DFT give a metal with no evidence for the Bi $6s$ charge disproportionation (all the Bi sites are found to have nearly same oxygen environment), implying the crucial importance of strong localization of the Ni $3d$ electrons due to correlation effects \cite{Subedi_2015}. 

Most importantly, our DFT+DMFT results provide a clear evidence that BiNiO$_3$ undergoes a structural transition from the triclinic insulating $P\bar{1}$ to orthorhombic metallic $Pbnm$ structure below ${\sim 0.95\,V_0}$ (above 8~GPa), in agreement with experiment \cite{Azuma2007,Carlsson2008,Azuma2011}.
We found that the transition pressure depends very sensitively on the choice of the Hubbard $U$ value, with $P_c \simeq 1$~GPa and 15~GPa for ${U=5}$~eV and $8$~eV, respectively.
The calculated bulk modulus ($U=6$~eV) is $K_0 \sim 143$~GPa, i.e., $K_0$ is found to decrease by $\sim$4\% upon the MIT into the metallic state. The latter is rather uncommon for a Mott MIT, indicating the importance of lattice effects at the MIT in BiNiO$_3$ \cite{Subedi_2015}.

The $Pbnm$ phase of BiNiO$_3$ is a correlated metal, characterized by a Fermi-liquid-like behavior with a weak damping of quasiparticles at the Fermi energy and by a substantial mass renormalization of $\frac{m^*}{m} \sim 2.5$ of the Ni $e_g$ bands. 
The Ni $e_g$ states show a quasiparticle peak at the Fermi level, with the upper Hubbard band at $\sim 1.0$~eV (see Fig.~\ref{fig:dos} and Supplementary Fig.~S3). 
The calculated Ni-ion local magnetic moment of ${1.3\,\mu_\textrm{B}}$ differs sufficiently from the fluctuating one ${\sim 0.5\,\mu_\textrm{B}}$, implying delocalization of the Ni $3d$ electrons at the transition. Indeed, our result for the local susceptibility shows itinerant-moment-like behavior, similar to that of the highly-pressurized $P\bar{1}$ phase (see Fig.~\ref{fig:chi_loc}).
The $Pbnm$ phase is found to be metallic for all studied here unit cell volumes, as well as even for a large Hubbard ${U=12}$~eV.
The pressure-induced MIT is found to be accompanied by a collapse of the lattice volume by ${\sim 5.2\%}$, resulting in the melting of charge disproportionation of the Bi sites. 
Thus, in the $Pbnm$ phase all the Bi sites are equivalent, whereas the Bi $6s$ states are fully occupied, i.e., Bi$^{3+}$. 
Moreover, our analysis of eigenvalues of the reduced Ni $3d$ density matrix suggests that the Ni sites are in a Ni$^{2+}$ state, with an atomic configuration $\simeq \sqrt{0.56}|d^8\rangle + \sqrt{0.30}|d^9\rangle$. We also notice a minor, below $\sim$10\%, contribution due to the $d^7$ atomic state, $\sqrt{0.09}|d^7\rangle$.
Based on this result, we conclude that no change of the valence state of the Ni$^{2+}$ ions occurs across the pressure-induced MIT in BiNiO$_3$, i.e., the Ni$^{2+}$ state remains stable.  
The latter is in a sharp contrast with the valence skipping Bi/Ni model proposed earlier for BiNiO$_3$~\cite{Azuma2007,Mizumaki2009}.
Our results suggest a novel microscopic mechanism of a Mott MIT under pressure
which is controlled by a charge-transfer between the $A$-site Bi $6s$ and ligand O $2p$ states. 
The pressure-induced MIT in BiNiO$_3$ is accompanied by a transition from the charge-disproportionated $\mathrm{Bi1}^{3+}_{0.5}(\mathrm{Bi2}^{(3 + \delta)+} \underline{L}^{2-\delta})_{0.5}$ to the charge-uniform
$\mathrm{Bi}^{3+} \underline{L}^{2}$ valence state.
The Bi $6s$ charge disproportionation (in the insulating $P\bar{1}$ phase) occurs together with the MIT, which follows rather than produces the structural transition.
We therefore conclude that the pressure-induced MIT and the concomitant melting of the Bi $6s$ charge ordering in BiNiO$_3$ is driven by the crystal structure transition. 
The latter highlights the complex interplay between the electronic structure and lattice effects in the vicinity of a Mott MIT in $R$NiO$_3$ nickelates~\cite{Subedi_2015}.
%

In conclusion, we employed the DFT+DMFT approach to determine the electronic structure and phase stability of paramagnetic BiNiO$_3$ across the pressure-induced Mott MIT. 
%
%
%
Our results for the $P\bar{1}$-structured BiNiO$_3$ under pressure propose a new mechanism for a correlation-driven metal-insulator transition, in which the Mott insulating state is (in much respect) controlled by the $s$-$p$ level splitting between the uncorrelated $A$-site Bi $6s$ and ligand O $2p$ states.
We show that the pressure-induced MIT in BiNiO$_3$ is associated with the melting of charge disproportionation of the Bi ions and is accompanied by delocalization of the Ni $3d$ electrons.
The phase transition results in a charge transfer between the Bi $6s$ and O $2p$ states, while the Ni sites remain to be Ni$^{2+}$.
Our results suggest that the pressure-induced change of the crystal structure drives the MIT in BiNiO$_3$. 
We argue that the $R$NiO$_3$ compounds (with $R=$rare earth and Bi) obey an intrinsic instability driven by the interplay of electron correlations and lattice effects, depending on the $R$-ion radius. It is associated with a crossover from charge disproportionation of the perovskite $B$-site Ni-ions (realized for the $R$-ions with the ionic radii smaller than that of La) to that of the $A$-site $R$-ions (for large $R$-ions), with LaNiO$_3$ being in between.   

We thank O. Peil for valuable discussions. 
We acknowledge
the support from Russian Foundation for Basic Research
(Project No. 18-32-20076).The DFT calculations were supported
by the Ministry of Science and Higher Education of
the Russian Federation (theme “Electron” No. AAAA-A18-
118020190098-5).


\pagebreak

\section*{Supplemental Material}
\beginsupplement

Under ambient pressure, BiNiO$_3$ adopts a highly distorted perovskite (triclinic) crystal structure with space group $P\bar{1}$ (see Supplementary Fig.~\ref{fig:structure}). It has two inequivalent Bi and four Ni sites and is characterized by the cooperative breathing Bi-O distortions of the lattice. The Bi sites are arranged in chains along the $c$ axis, with a checkerboard pattern in the $ab$-plane. The $P\bar{1}$-structured BiNiO$_3$ is an insulator with an energy gap of $\sim 0.68$ eV as estimated from the electrical resistivity \cite{Ishiwata2002}.

\begin{figure*}[h]
\includegraphics[trim=0cm 0cm 0cm 0cm,width=0.66\textwidth]{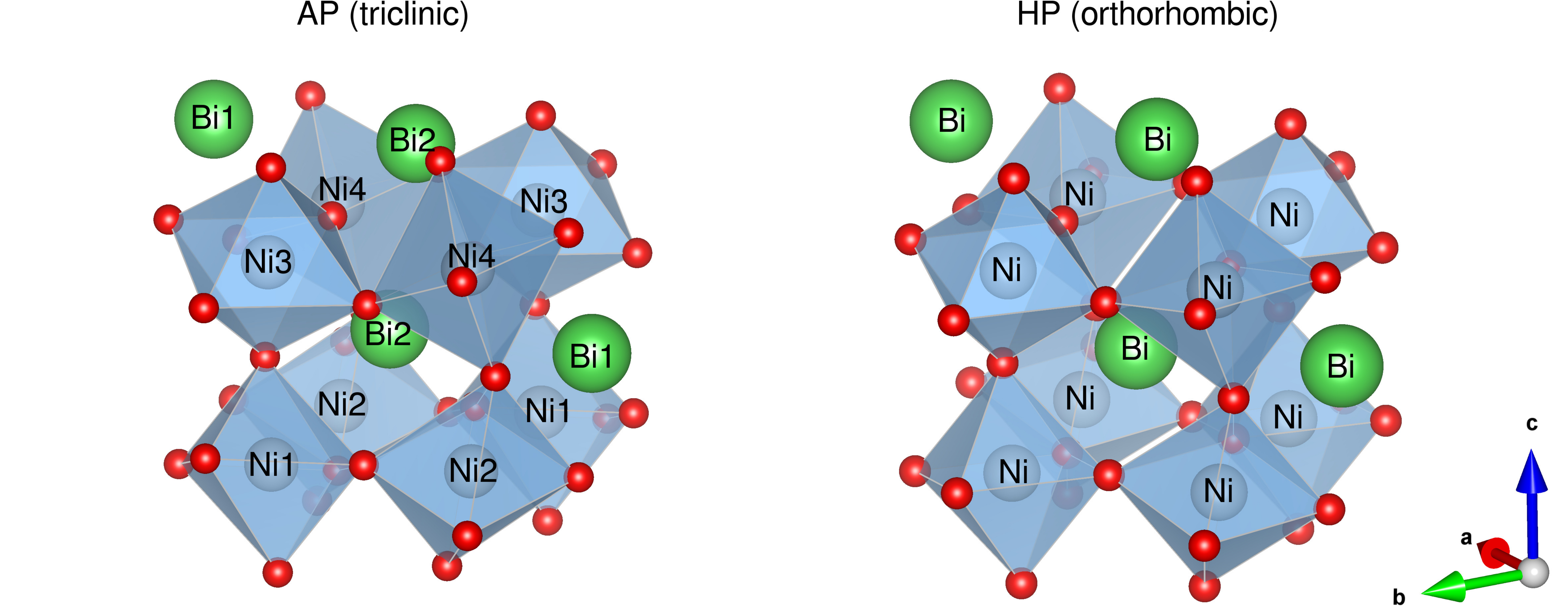}
\caption{(Color online)
Crystal structure of the triclinic $P\bar{1}$ (left) and highly-pressurized orthorhombic $Pbnm$ (right) phases of BiNiO$_3$. The oxygen atoms are depicted by small red balls. The figure was prepared with the VESTA program~\cite{vesta}.}
\label{fig:structure}
\end{figure*}

Under pressure above $\sim4$~GPa, BiNiO$_3$ undergoes a structural transformation to the orthorhombic GdFeO$_3$-type ($Pbnm$) crystal structure, which has a single type of Bi and Ni ions. The phase transition is accompanied by a Mott insulator-to-metal transition and is associated with suppression of the breathing distortions of the lattice (all the Ni and Bi sites become equivalent in the $Pbnm$ phase).

\begin{figure*}[h]
\begin{tabular}{cc}
\includegraphics[trim=0cm 0cm 0cm 0cm,width=0.41\textwidth]{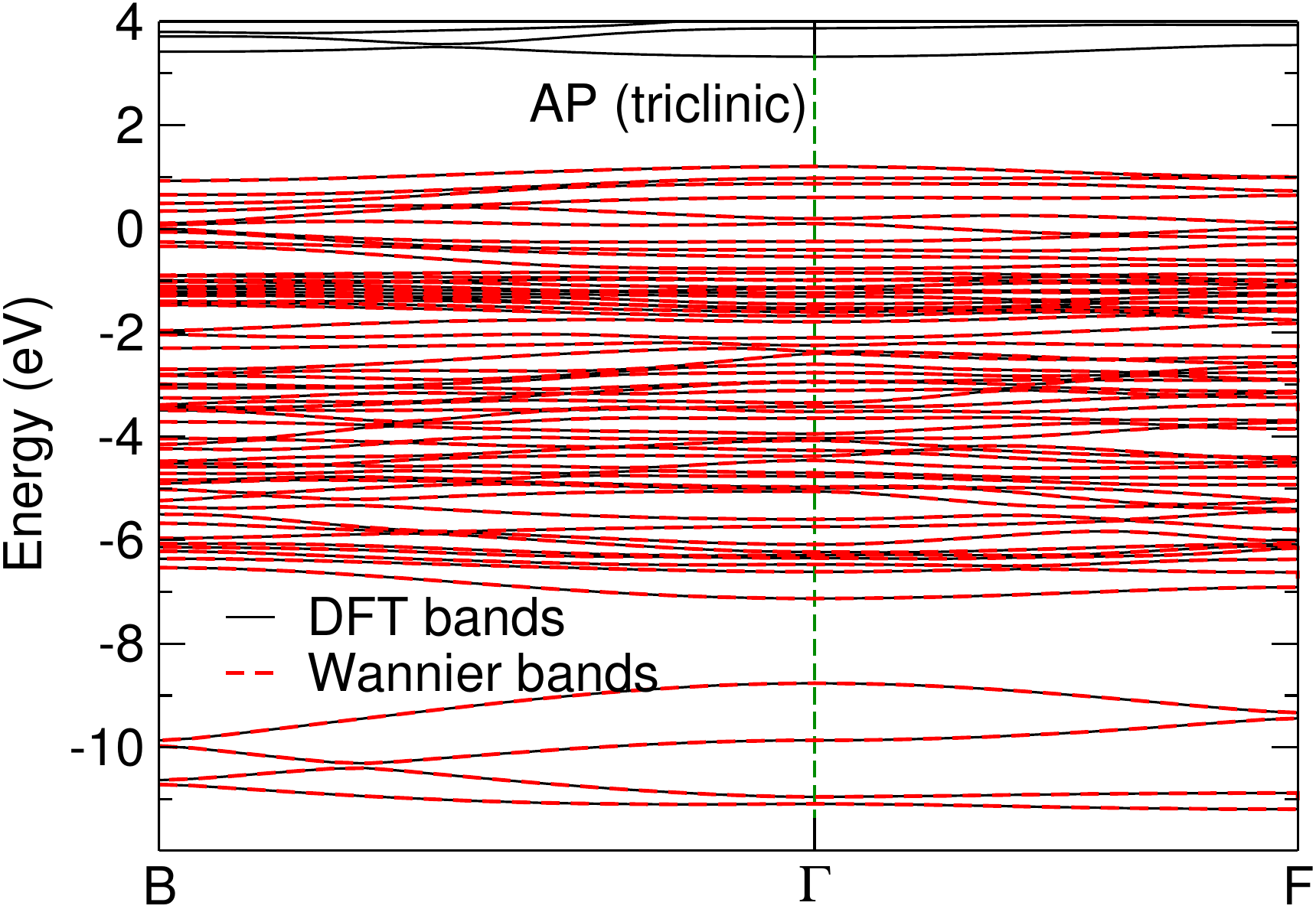}
\includegraphics[trim=0cm 0cm 0cm 0cm,width=0.41\textwidth]{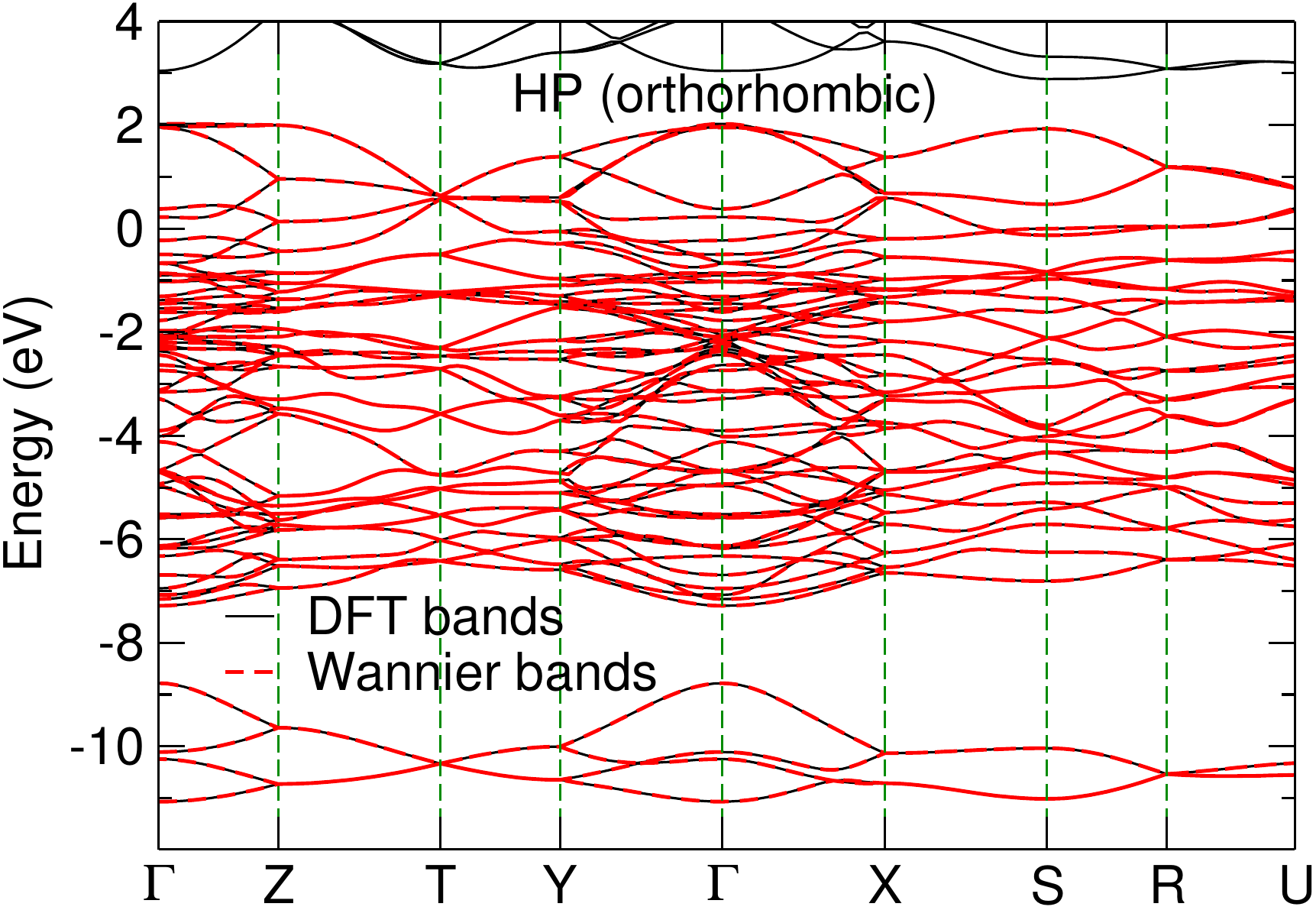}
\end{tabular}
\caption{(Color online).
Band structure of BiNiO$_3$ calculated within nonmagnetic DFT for the ambient-pressure $P\bar{1}$ (left panel) and high-pressure $Pbnm$ (right panel) phases
in comparison with the Wannier bands corresponding to the constructed Bi $6s$, Ni $3d$, and O $2p$ Wannier functions (red dashed lines).
The Fermi level is at zero energy.
\label{fig:band_structure}}
\end{figure*}

Here, we employed the DFT+DMFT approach to explore the electronic properties and phase stability of paramagnetic BiNiO$_3$ under pressure 
using the DFT+DMFT method \cite{dftdmft} implemented with plane-wave pseudopotentials \cite{espresso,Leonov1}.
We start by constructing the effective low-energy Hamiltonian [$\hat{H}^{\mathrm{DFT}}_{\sigma,\alpha\beta}(\bf{k})$], which explicitly contains the Bi $6s$, Ni $3d$, and O $2p$ valence states, using the projection onto Wannier functions~\cite{Wannier1}. For this purpose, for the partially filled Bi $6s$, Ni $3d$, and O $2p$ orbitals we construct a basis set of atomic-centered symmetry-constrained Wannier functions \cite{Wannier2}.
The Wannier functions are constructed over the full energy range spanned by the $s$-$p$-$d$ band complex using the scheme of Ref.~\cite{Wannier2}.
We obtain the $s$-$p$-$d$ Hubbard Hamiltonian (in the density-density approximation)
\begin{eqnarray}
\label{eq:hamilt}
\hat{H} = \sum_{\bf{k},\sigma} \hat{H}^{\mathrm{DFT}}_{\sigma,\alpha\beta}({\bf{k}}) + \frac{1}{2} \sum_{i,\sigma\sigma',\alpha\beta} U_{\alpha\beta}^{\sigma\sigma'} \hat{n}_{i,\alpha\sigma} \hat{n}_{i,\beta\sigma'} - \hat{H}_{\mathrm{DC}},
\nonumber
\end{eqnarray}
where $\hat{n}_{i,\alpha\sigma}$ is the occupation number operator for the $i$-th Ni site with spin $\sigma$ and (diagonal) orbital indices $\alpha$.
In Supplementary Fig.~\ref{fig:band_structure} we show our results for the band structure of BiNiO$_3$ calculated within nonmagnetic DFT in comparison with the Wannier Bi $6s$, Ni $3d$, and O $2p$ band structure for the 
ambient-pressure $P\bar{1}$ and high-pressure $Pbmn$ phases of BiNiO$_3$. Our results for the leading Wannier hopping integrals between the Bi $6s$ and neighbor ions in the ambient-pressure $P\bar{1}$ and high-pressure $Pbnm$ phases of BiNiO$_3$ are summarized in Table~\ref{tab:model_1}.
All the calculations are performed in the local basis set determined by diagonalization of the corresponding Ni $3d$ occupation matrices. 

In order to solve the realistic many-body problem, we employ the continuous-time hybridization-expansion quantum Monte-Carlo algorithm \cite{CT-QMC}. The Coulomb interaction has been treated in the density-density approximation. The elements of the $U$ matrix are parametrized by the average Coulomb interaction $U$ and Hund's exchange $J$ for the Ni $3d$ shell. For all the structural phases considered here we have used the same $U=6$ eV and $J=0.95$~eV values as was estimated previously for $R$NiO$_3$ \cite{Park2012,Nowadnick2015}. 
The spin-orbit coupling was neglected in these calculations. Moreover, the $U$ and $J$ values are assumed to remain constant upon variation of the lattice volume.
We employ the fully localized double-counting correction, evaluated from the self-consistently determined local occupations, to account for the electronic interactions already described by DFT,
$\hat{H}_{DC}=U ( N - \frac{1}{2} ) - J ( N_{\sigma} - \frac{1}{2} )$,
where $N_\sigma$ is the total Ni $3d$ occupation with spin $\sigma$ and $N=N_\uparrow+N_\downarrow$.
Here, we employ a fully self-consistent in charge density DFT+DMFT scheme in order to take into account the effect of charge redistribution caused by electronic correlations and electron-lattice coupling.

\begin{figure*}[h]
\begin{tabular}{cc}
\includegraphics[trim=0cm 0cm 0cm 0cm,width=0.41\textwidth]{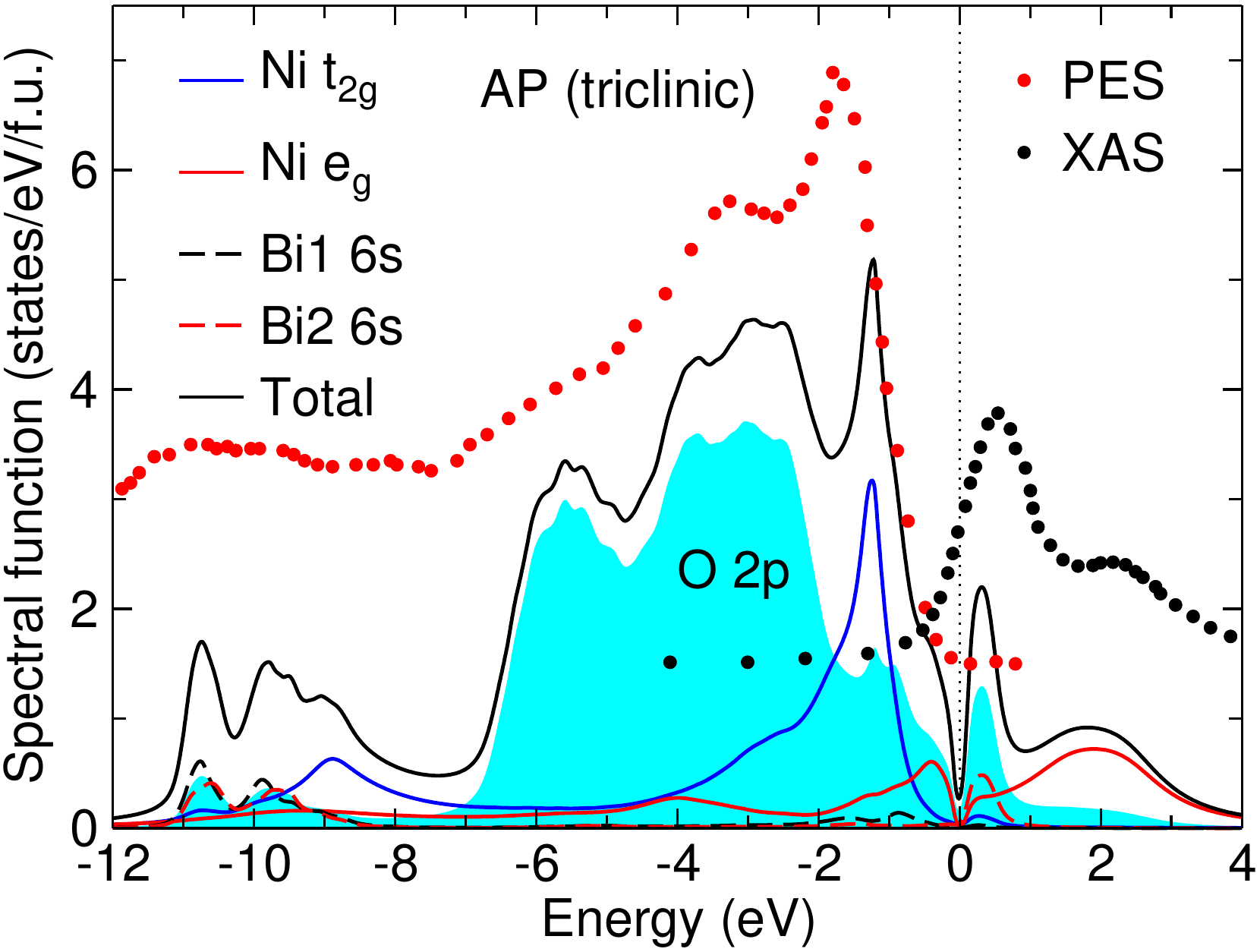}
\includegraphics[trim=0cm 0cm 0cm 0cm,width=0.41\textwidth]{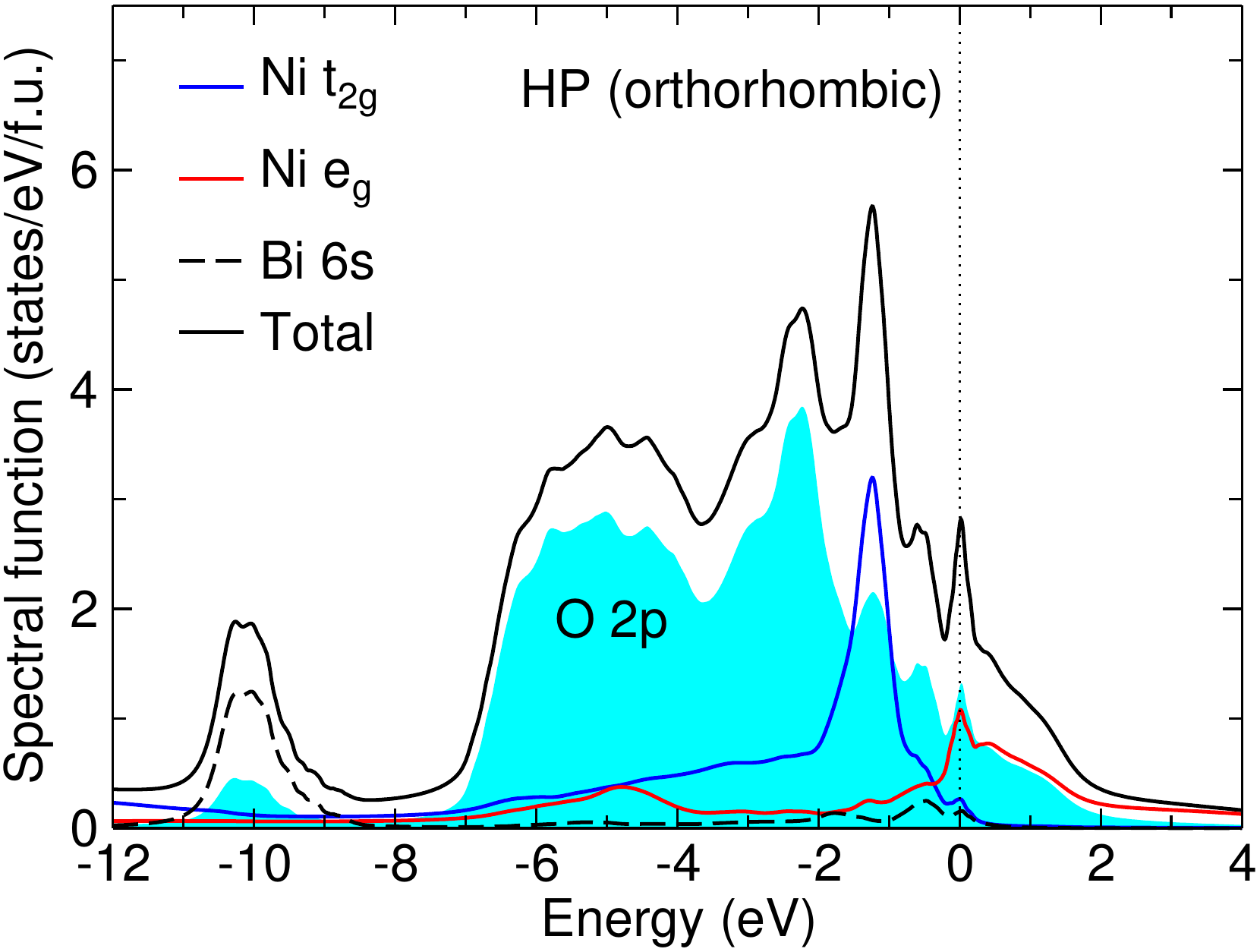}
\end{tabular}
\caption{(Color online).
Orbitally-resolved spectral functions of paramagnetic BiNiO$_3$ calculated within DFT+DMFT for the ambient-pressure $P\bar{1}$ (left panel) and high-pressure $Pbnm$ (right panel) phases of BiNiO$_3$. Photoemission (PES) and X-ray absorption (XAS) spectra are shown for comparison~\cite{Wadati2005}.
The DFT+DMFT calculations are performed at a temperature ${T = 387}$~K (above $T_\mathrm{N} \sim 300$~K). The Fermi level is at zero energy.
\label{fig:dos_pade}}
\end{figure*}

In Supplementary Fig.~\ref{fig:dos_pade} we show the spectral functions of paramagnetic BiNiO$_3$ calculated by DFT+DMFT in comparison with photoemission (PES) and X-ray absorption (XAS) spectra taken at room temperature~\cite{Wadati2005}.
Our calculations are performed in the paramagnetic state at a temperature ${T = 387}$~K, above the N\'eel temperature ${T_\mathrm{N}\sim 300}$~K.
To calculate the spectral functions, we employ the Pad\'{e} analytical continuation procedure for the self-energy.
In our calculations we adopt the experimental crystal structure data (atomic positions for the orthorhombic phase are taken from the experiment at a pressure of $\sim$7.7~GPa~\cite{Azuma2007}).

The calculated spectral functions are in overall good agreement with the experimental spectra.  
In particular, in the insulating triclinic phase, the energy gap lies between the occupied and unoccupied Ni $e_g$ states, strongly mixed with the O $2p$ and the empty Bi2 $6s$ states (the Bi1 $6s$ states are fully occupied).
Our results indicate that all the Ni sites (the insulating $P\bar{1}$ phase has four inequivalent Ni sites) are nearly equivalent.
A sharp peak at about -1.5~eV originates from the occupied Ni $t_{2g}$ states, which form a lower Hubbard band at -9~eV.
The PES spectral weight lying at about -3 and -5~eV is mainly due to the O $2p$ states, the hump at -10~eV is predominantly due to the Bi $6s$ states.
In the metallic orthorhombic phase, the peak at the Fermi level and the spectral weight at the bottom of the conduction band are predominantly formed by the Ni $e_g$ and O $2p$ states. The Ni $e_g$ upper Hubbard band appears at $\sim 1.0$~eV.
The peak at about -1.5~eV is due to the occupied Ni $t_{2g}$ states.
In contrast to the insulating phase, all the Bi states are occupied and are located at about -10~eV.

\begin{table*}[t]
  \caption{Leading Wannier hopping integrals (in meV) between Bi $6s$ and neighbor ions in the ambient-pressure $P\bar{1}$ (left part) and high-pressure $Pbnm$ (right part) phases of BiNiO$_3$.
  \label{tab:model_1}}
  \begin{ruledtabular}
  \begin{tabular}{cc} 
    \begin{tabular}{cccc}
      Atom \quad &   Atom \quad &  Distance (a.u.) \quad &  Hoppings (meV)     \\
\cline{1-4}
        Bi $6s$  &  O $2p$      &  4.08   &  -1304, -1234, -71  \\
        Bi $6s$  &  O $2p$      &  4.11   &  -1410,  1037,  631  \\
        Bi $6s$  &  O $2p$      &  4.48   &  -280, -404,  1086  \\
        Bi $6s$  &  O $2p$      &  4.62   &   772,  295,  1144  \\
        Bi $6s$  &  O $2p$      &  4.86   &  -674,  72,  839  \\
        Bi $6s$  &  O $2p$      &  4.90   &   422,  47, -916  \\
        Bi $6s$  &  O $2p$      &  5.41   &   052, -80, -757  \\
        Bi $6s$  &  O $2p$      &  5.98   &  -385,  78,  204  \\
        Bi $6s$  &  Ni $e_g$    &  6.03   &   41,  2  \\ 
        Bi $6s$  &  Ni $t_{2g}$ &  6.03   &  -37,  75,  163  \\
        Bi $6s$  &  Ni $e_g$    &  6.10   &  -7,  58  \\
        Bi $6s$  &  Ni $t_{2g}$ &  6.10   &   31, -126, -142  \\
        Bi $6s$  &  Ni $e_g$    &  6.11   &   11,  48 \\
        Bi $6s$  &  Ni $t_{2g}$ &  6.11   &  -2, -274, -163  \\
        Bi $6s$  &  Ni $e_g$    &  6.16   &   40,  12 \\
        Bi $6s$  &  Ni $t_{2g}$ &  6.16   &   58, -14,  64 \\
    \end{tabular}
    &
%
    \begin{tabular}{cccc}
      Atom \quad &   Atom \quad &  Distance (a.u.) \quad &  Hoppings (meV)     \\
\cline{1-4}
        Bi $6s$  &  O $2p$      &  4.24   &  0,  1709, 242   \\
        Bi $6s$  &  O $2p$      &  4.44   & -271, -1536, 95  \\
        Bi $6s$  &  O $2p$      &  4.44   & -271, -1536, -95  \\
        Bi $6s$  &  O $2p$      &  4.58   &  0,  622,  1132   \\
        Bi $6s$  &  O $2p$      &  4.85   & -69, -103,  957   \\
        Bi $6s$  &  O $2p$      &  4.85   & -69, -103,  -957   \\
        Bi $6s$  &  O $2p$      &  4.95   &  6, -142,  935    \\
        Bi $6s$  &  O $2p$      &  4.95   &  6, -142, -935    \\
        Bi $6s$  &  O $2p$      &  5.85   &  0, -298, -151    \\
        Bi $6s$  &  O $2p$      &  5.85   &  0, -49, -330     \\
        Bi $6s$  &  Ni $e_g$    &  5.86   & -6, -11         \\ 
        Bi $6s$  &  Ni $t_{2g}$ &  5.86   & 40, -12,  227   \\
        Bi $6s$  &  Ni $e_g$    &  5.86   &  6,  11         \\ 
        Bi $6s$  &  Ni $t_{2g}$ &  5.86   & 40, -12,  227   \\
        Bi $6s$  &  Ni $e_g$    &  6.11   &  0, -40         \\
        Bi $6s$  &  Ni $t_{2g}$ &  6.11   &  38,  156, -55  \\
    \end{tabular}
   \end{tabular}
 \end{ruledtabular}
\end{table*}

\end{document}